\documentclass[fleqn,usenatbib]{mnras}

\usepackage{newtxtext,newtxmath}
\usepackage[T1]{fontenc}
\usepackage{graphicx}	
\usepackage{amsmath}	

\title[On the origin of Mixed Morphology SNRs]{On the origin of mixed morphology supernova remnants: Linking their properties to the evolution of a red supergiant progenitor star
}

\author[Chiotellis,  Zapartas \& Meyer]{
Alexandros Chiotellis,$^1$\thanks{E-mail: a.chiotellis@noa.gr}
Emmanouil Zapartas $^{2,1}$
and Dominique M.-A. Meyer $^3$
\\
$^{1}$ Institute for Astronomy, Astrophysics, Space Applications
and Remote Sensing, National Observatory of Athens,
15236 Penteli, Greece \\
$^{2}$ Institute of Astrophysics, FORTH, N. Plastira 100,  Heraklion, 70013, Greece \\
$^{3}$ Institute of Space Sciences (ICE, CSIC), Campus UAB, Carrer de Can Magrans s/n, 08193 Barcelona, Spain \\
}

\date{Accepted 27 March 2024} 

\pubyear{2024}

\begin{document}
\label{firstpage}
\pagerange{\pageref{firstpage}--\pageref{lastpage}}
\maketitle

\begin{abstract}
Mixed-morphology supernova remnants (MMSNRs) are characterized by  a shell-like morphology in the radio and centrally-peaked thermal emission in the X-ray band. The nature of this peculiar class of supernova remnants (SNRs) remains a controversial issue. In this work, by pairing the predictions of  stellar evolution theory with two-dimensional hydrodynamic simulations, we show that the mixed morphology properties of a SNR can arise by the interaction of the SNR with the circumstellar medium shaped by a red supergiant  progenitor star, embedded in a dense environment. As a study case, we model the circumstellar medium formation and the subsequent interaction of the SNR with it of a $15~\rm M_{\odot}$ progenitor star. The reflected shock, formed by the collision of the SNR with the density walls of the surrounding circumstellar cavity, accumulates and re-shocks the supernova ejecta at the center of  the remnant, increasing its temperature so that the gas becomes X-ray bright. Such a formation mechanism may naturally explain the nature of MMSNRs resulted from Type II supernovae without the demand of additional physical mechanisms and/or ambient medium inhomogeneities. We discuss alternative evolutionary paths that potentially could be ascribed for the MMSNR formation within the framework of the reflected shock model.

\end{abstract}




\section{Introduction}

Mixed morphology or thermal composite supernova remnants (hereafter MMSNRs) consist a distinctive and peculiar class of supernova remnants (SNRs) that up to date counts more than 37 members \citep{Vink2012,Zhang2015}. MMSNRs are characterized by the coexistence of a shell-like radio morphology and centrally peaked, thermal X-ray emission \citep{Rho1998,Jones1998J, Wilner1998}. They tend to be evolved SNRs ($t \approx 10^4$~yr)  and are met in the denser parts of the interstellar medium (ISM). The majority of MMSNRs are associated with OH masers \citep{Green1997,Yusef2003} something that indicates interaction of the SNR with surrounding molecular clouds \citep[e.g.][]{Claussen1997,Frail1998,Arias2019}.  Regarding the properties of their thermal emission, X-ray spectroscopy reveals that several cases of MMSNRs show evidence of overionization \citep[e.g.][]{Kawasaki2005}  while the central X-ray emitting gas is characterized by enhanced chemical abundances \citep{Lazendic2006, Bocchino2009, Pannuti2014}.

The morphology of  MMSNRs is puzzling as it does not obey  the standard SNR evolution models.  The large angular size of these remnants, their radio shell emission, and the bright optical filaments that these remnants display, advocate that MMSNRs are mature, slowly expanding SNRs,  being evolved beyond the Sedov-Taylor phase. Such a conclusion is in sharp contrast with the central X-ray emission of MMSNRs, given that evolved SNRs are expected to be characterized by a low density and cold interior \citep{ Vink2012}. In addition, the fact that in MMSNRs  there is no obvious external source -such as an active pulsar- responsible for their central X-ray emission makes their nature even more puzzling. 

Several possible mechanisms have been proposed aiming to explain the peculiar properties of MMSNRs,  such as a radiatively cooled rim \citep{Harrus1997, Rho1998}, thermal conduction in the interior hot gas \citep{Cox1999, Shelton1999}, evaporation of gas from the shock-engulfed cloudlets \citep{White1991, Slavin2017, Zhang2019, Okon2020} and even projection effects for some particular SNRs \citep{Petruk2001}.

\citet{Chen2008} in order to explain the double X-ray emitting shells observed at the interior of the MMSNR Kesteven~27,  first suggested the reflected shock scenario. According to it, the collision of the SNR with the density walls of a pre-existing cavity, formed by the stellar progenitor wind, triggered the formation of a reflected shock that reheated the SN ejecta and shaped the X-ray shells of the remnant. The same scenario was suggested for the MMSNRs Kesteven~41 \citep{Zhang2015} and HB3 \citep{Boumis2022}, for which the authors claimed that the remnants are interacting with the cavity walls formed by a  $\sim 18 ~\rm M_{\odot}$ and $\sim 34 ~\rm M_{\odot}$ progenitor star, respectively.     Nevertheless, \citet{Dwarkadas2013} tested the idea of \citet{Chen2008} and by conducting hydrodynamic simulations they attempted to reproduce the  properties of Kesteven~27. The authors concluded that the collision of a SNR with a cavity's  wall is not able to reproduce the mixed morphology properties of the remnant as the formed reflected shock propagates into a low-density ejecta and consequently the resulting central X-ray flux is very faint.  On the other hand, reflected shock models achieved to reproduce observed properties of specific MMSNRs but under the consideration of very specific circumstellar medium (CSM) and/or ISM conditions, different than the wind-blown cavity and additionally employing the process of thermal conduction. In particular, \citet{Zhou2011} modeled the MMSNR W49B  assuming that the explosion center was surrounded by a dense ring and a dense cloud at its northern region, while \citet{Ustamujic2021} modeled the MMSNR IC 443 considering  a toroidal molecular could and a spherical cap of dense atomic gas on the top, lying around the progenitor system.

  In this work, we re-assess the reflected shock model for MMSNRs evolving within the wind-blown cavity formed by the progenitor star by taking into account all the   outcomes of stellar evolution for Type~II supernovae (SNe), i.e. SNe resulted by Red Supergiant (RSG) stars.  By coupling stellar evolution models with two-dimensional (2D) hydrodynamic simulations we show that the mixed morphology properties of SNRs can naturally arise by the interaction of the SN ejecta with the circumstellar structure shaped by the progenitor star embedded in a dense ISM, without
any additional demand of a specific physical process (e.g. thermal conduction) or peculiar properties of the ambient medium (e.g. dense blobs, rings or tori).

This paper is organised as follows: In Section \ref{Sect2} we discuss the properties of MMSNRs progenitor stars as extracted by the relevant observations. In Section \ref{Sect3} we describe the followed methodology of our modeling and the initial set up of the simulations. The results of our modelling and their analysis are presented in Section \ref{Sect4}. Finally, we sum up and we discuss our main conclusions in Sections \ref{Sect5}.

\begin{figure}
\includegraphics[trim=30 30 10 10 10,clip=true,width=\columnwidth,angle=0]{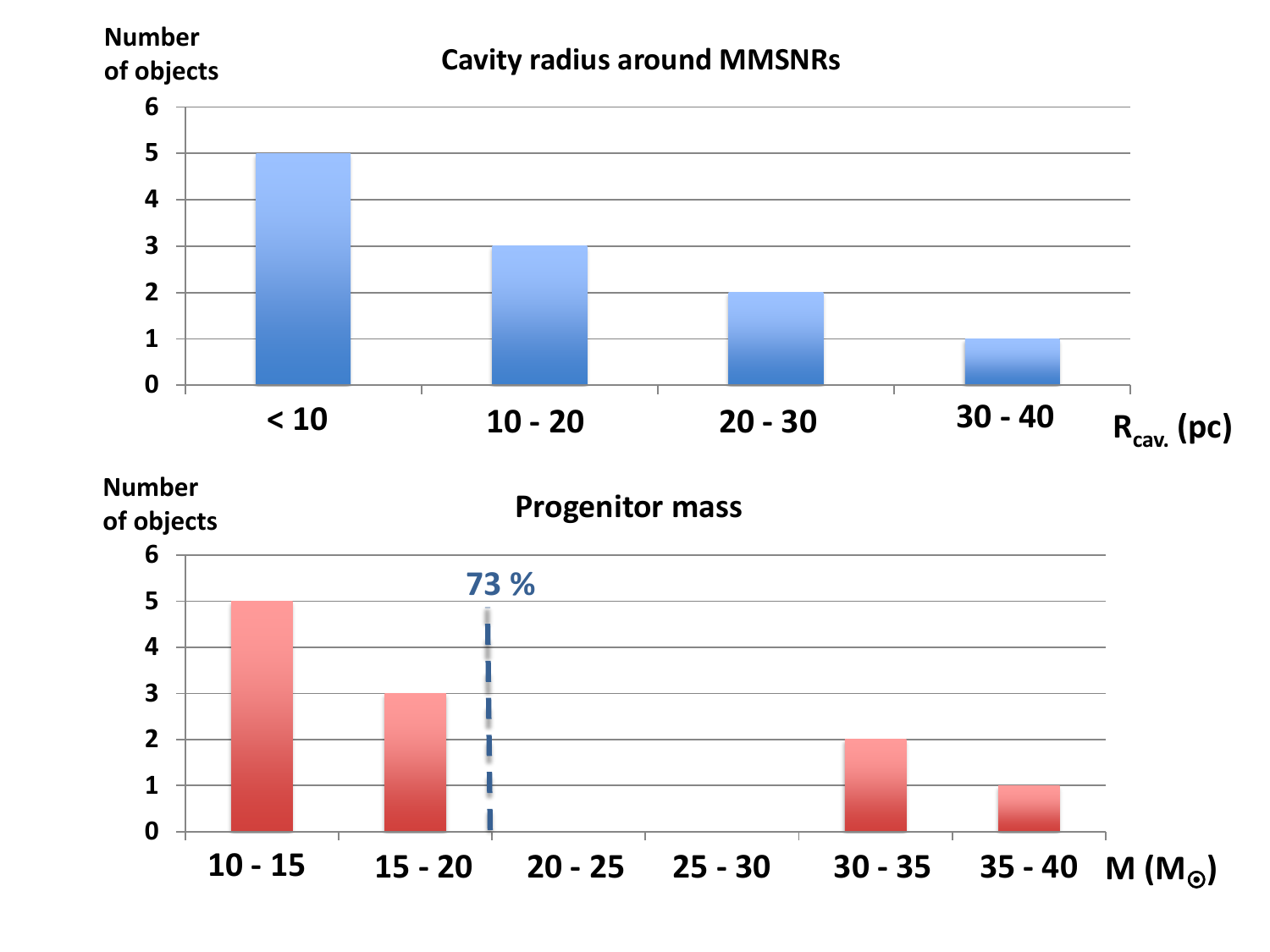}
\caption {Upper plot: The size distribution of the observed cavities around MMSNRs as extracted by Table 4 of \citet{Zhang2015}. Lower plot: The corresponding distribution of the progenitor's mass resulted by the wind-blown cavity size- stellar mass relation of \citet{Chen2013}. 
}
\label{fig:histrogram}
\end{figure}

\section{Stellar progenitors of Mixed Morphology Supernova Remnants}
\label{Sect2}

Determining the stellar progenitor of evolved SNRs is a difficult task as they do not contain an active reverse shock  and thus, any information on the ejecta's chemical composition has been lost. A promising method through which we gain indirect evidence on the nature and evolution of their parent stellar systems is the study of  the surrounding CSM \citep[e.g.][]{Tsebrenko2013, Chiotellis2013, Chiotellis2021, Orlando2022, meyer_mnras_493_2020,  Meyer2022, Meyer2023, Velazquez2023, Villagran2024}.

Several MMSNRs have been observed to evolve into a cavity sculptured by the wind of the progenitor star. For eleven cases the size of this cavity has been adequately estimated \citep[see table 4 of ][and references therein]{Zhang2015}, the distribution of which is illustrated in the upper histogram of Figure \ref{fig:histrogram}. Employing in these observational data the linear relationship for wind-blown cavities sizes of main-sequence OB stars of \citet{Chen2013} ($R_b \approx 1.22~{~\rm M/M_{\odot}}-9.16$~pc) we get the corresponding progenitor's initial mass distribution (lower plot of Fig. \ref{fig:histrogram}) \footnote{For high mass stars ($M> 20 \rm M_{\odot}$) the \citet{Chen2013} relationship is not accurate as it takes into account only the main-sequence wind and not the subsequent Wolf-Rayet wind.}.  The mass distribution of the MMSNR progenitors follows an exponential decay where the majority ($\sim 73 \%)$ of the systems have a mass between $10 - 20~\rm M_{\odot}$.

Stars in this mass range will  evolve as RSGs before their final explosion as Type II SNe \citep{Heger+2003}. Thus, the cavity sculptured by the fast main-sequence wind is expected to be partially filled by a dense wind bubble formed by the slow and strong RSG wind. Hence, when the SN explosion will occur, the stellar ejecta will first interact with the RSG wind bubble before its collision with the density walls of the surrounding cavity. The aim of this work is to evaluate the effect of this interaction on the resulting properties of the MMSNR, a process that in the previous models had not been taken into account. The adopted methodology to achieve this is described in the following section.

\begin{figure}
\includegraphics[trim= 0 440 0 0, width=\linewidth,angle=0]{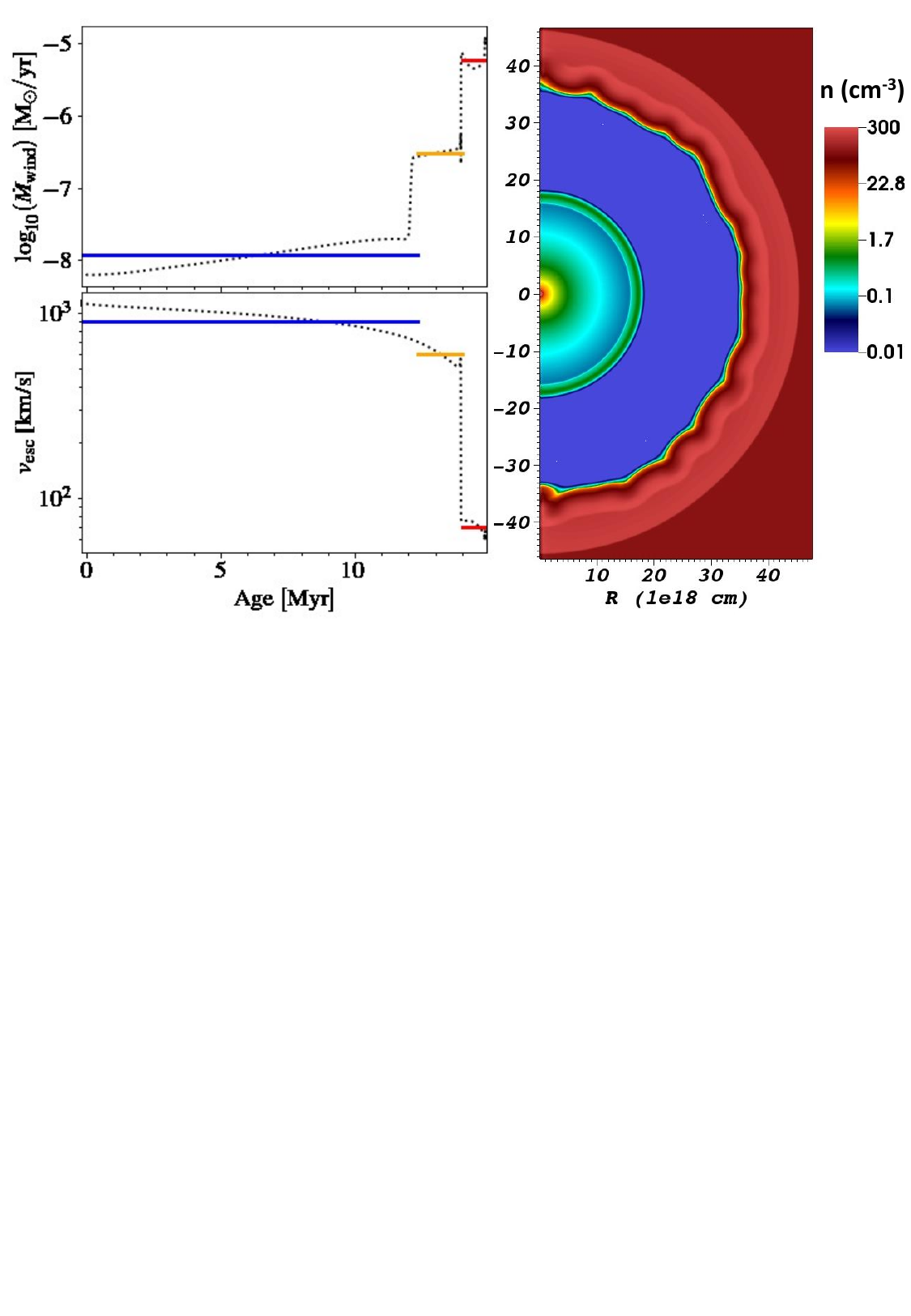}
\caption {{\it Left panels:} Evolution over time of the mass-loss rate (top) and surface escape velocity  (bottom) of a $15~\rm M_{\odot}$  {\sc posydon} stellar progenitor, simulated with {\sc mesa} code. The blue/orange/red horizontal lines depict the values assumed during the early MS / late MS / RSG phase, respectively (summarized in Table~{\ref{tab:wind}}). {\it Right panel:} The resulting CSM geometry  after applying these wind mass loss assumptions to an ambient ISM density of $n_{\rm ism} = 100~\rm cm^{-3}$.}
\label{fig:progenitor}
\end{figure}

\begin{table*}
\caption{Summary of the assumed wind mass loss rate and velocity of the SN progenitor, at different evolutionary phases
\label{tab:wind}}.
    \centering
    \begin{tabular}{c||c|c|c|c|c}
       
            Wind  &  Mass-loss & Duration [Myr] &  $v_{\rm esc}$ & Multiplication &  $v_{\rm wind}$ \\
         Phase  &  rate [ $\rm {M_{\odot}~yr^{-1}}$]  & (= End - Beginning of phase) & [$\rm km~s^{-1}$] & factor  &  [$\rm km~s^{-1}$] \\
         \hline
        Early MS & $1.2*10^{-8}$ & $12.4 \, (=12.4-0)$ & 900 & 2.6 & 2340\\
        Late MS & $3*10^{-7}$ & $1.6 \, (=14.0 - 12.4)$ & 600 & 1.3 & 780\\
        RSG phase & $6*10^{-6}$ & $0.9 \, (=14.9 - 14)$ & 70 & 0.30 & 20\\ 
    \end{tabular}
\end{table*}

\section{Method} \label{Sect3}

We perform 2D hydrodynamic simulations employing the hydrodynamic code {\sc amrvac} \citep{Keppens03}. The models were carried out on   spherical coordinates ($R,\theta$) assuming symmetry in the
third dimension of the azimuthal angle ($\phi$). The radial span of the computational domain is $R= 6.4 \times 10^{19}$~cm, while the polar angle $\theta$ ranges from $\rm 0^o$ to $\rm 180^o$.  Our grid is divided into $(R \times \theta) = 360 \times 120$ grid cells.  We additionally employ  the adaptive mesh capabilities of the {\sc amrvac} code by using three refinement levels of resolution, at each of which the resolution
is doubled as a result of large gradients in density and/or energy.
Hence, the maximum effective resolution becomes $4.4 \times 10^{15}$ cm
by $\rm 0.37^o$. Radiative cooling is prescribed using the cooling curve of \citet{Schure09}.

We conduct our simulations in two steps. We first simulate the formation of the CSM bubble formed by the mass outflows of the stellar progenitor. As a case study, we consider a progenitor star with an initial mass of 15~$\rm{M_{\odot}}$ which falls within the typical masses inferred for observed MMSNR (Figure~\ref{fig:histrogram}). The stellar wind properties were extracted by a non-rotating, single star model of 15~$\rm{M_{\odot}}$ at solar metallicity, simulated by {\sc mesa} stellar evolution code \citep{Paxton+2011, Paxton+2013, Paxton+2015, Paxton+2018, Paxton+2019} within the framework and according to the assumptions in   {\sc posydon}  \citep{Fragos+2023}. In the hydrodynamic simulations, the stellar wind is imposed in the inner radial boundary of the grid in the form of a continuous inflow with a density profile  $\rho=\dot{M}_w / (4\pi u_w r^2 )$ and  momentum  $m_r = \rho u_w$ and $m_{\theta} = 0$. The ambient ISM is  considered to be homogeneous with a density of $n_{\rm ism}= 100~\rm cm^{-3}$ and temperature  $T_{\rm ism}= 100$~K, representative of the dense ISM that the majority of MMSNRs are met \citep{Vink2012}.

In the second step of our hydrosimulations, we introduce at the center of the formed circumstellar wind bubble the supernova ejecta with mass $M_{\rm ej}= 8.0 ~ \rm M_{\odot}$, derived by the final, pre-SN stellar mass of the progenitor which is $\sim 9.4~{\rm M_{\odot}}$ and assuming that 1.4~${\rm M_{\odot}}$  collapsed to form a typical neutron star.  We also assume a typical SN  energy of $E_{\rm ej}=10^{51}~\rm erg$  and we let the SN to evolve and interact with the surrounding CSM. Following \cite{Truelove99}, the SN ejecta density profile is described by a constant density core with an envelope that follows a power law of $\rho \propto r^{-n}$ with n = 9, while the ejecta’s velocity increases linearly.

\begin{figure*}
\includegraphics[trim=0 150 0 32 ,clip=true,width=\linewidth,angle=0]{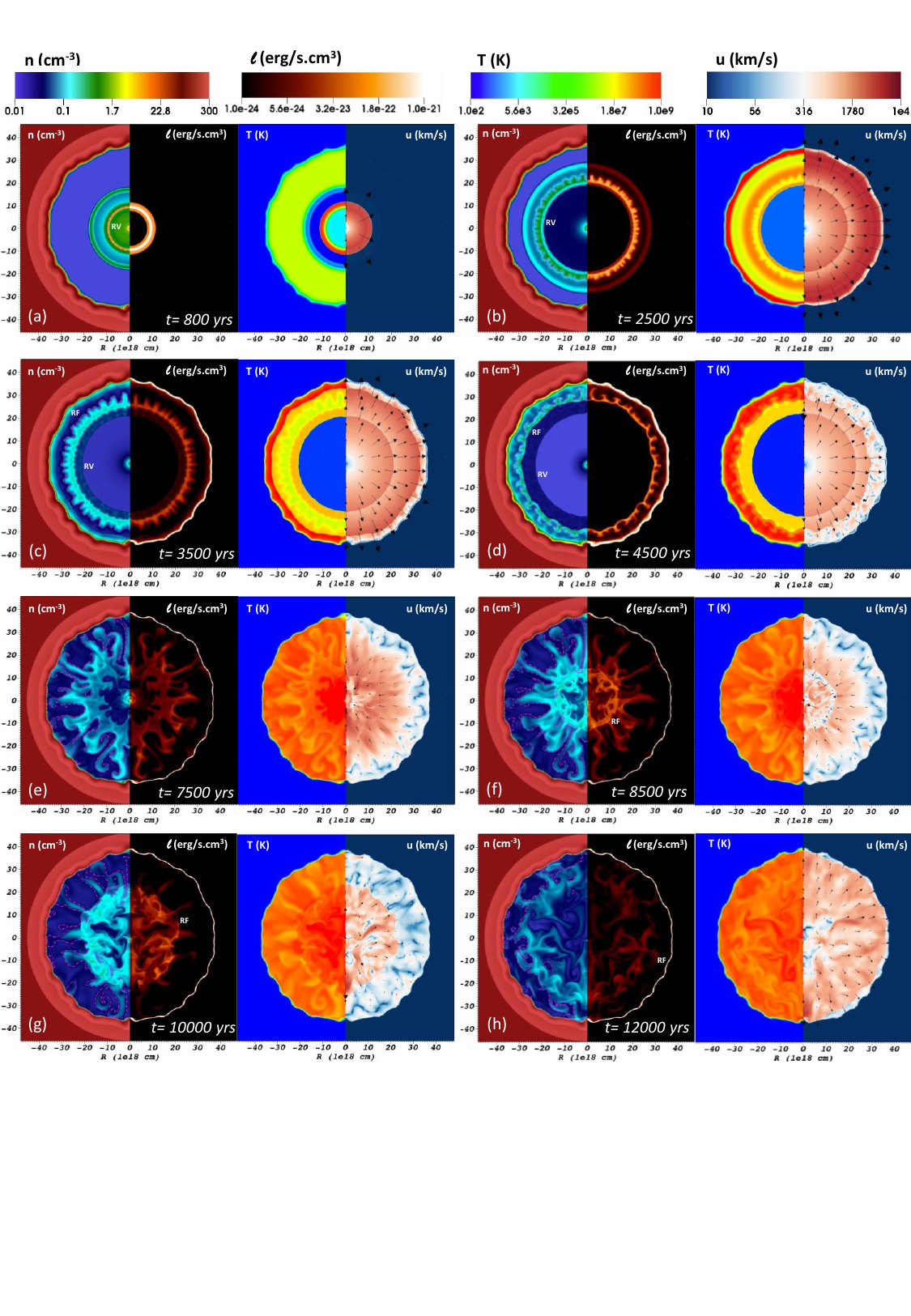}
\caption {The 2D  maps of the SNR evolution for eight different snapshots. The color bars - from left to right- refer to gas':  number density, luminosity per unit volume, temperature and velocity. The dotted pink line at the number density plots points the border between the ejecta and the CSM dominated material, while the black arrows at the velocity plots indicate the gas direction of motion. The symbols `RV' and `RF' refer to the position of the reverse shock and reflected shock, respectively.  }
\label{fig:MMSNR}
\end{figure*}

\begin{figure*}
\includegraphics[trim=0 440 0 5 ,clip=true,width=\linewidth,angle=0]{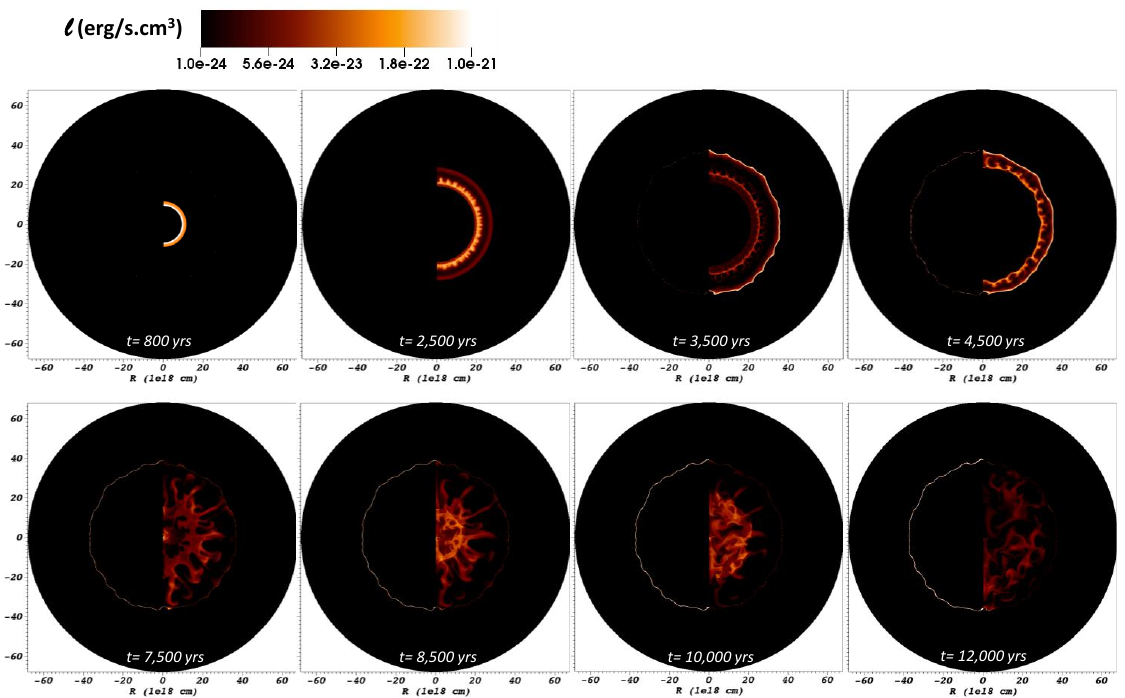}
\caption { The 2D  maps of the specific luminosity (luminosity per unit volume)  for the eight different snapshots of the SNR evolution depicted in Fig. \ref{fig:MMSNR}. In each panel the left part of the circle corresponds to the specific luminosity of the gas with temperature lower than $10^4$~K and the right one to those with temperature higher than $10^7$~K (see text for details).}
\label{fig:Xray_opt}
\end{figure*}

\section{Results} \label{Sect4}

\subsection{Stellar wind and CSM properties}

The progenitor's wind properties are summarized in Table~\ref{tab:wind}. The implemented wind prescriptions of the $\rm 15~\rm M_{\odot}$ stellar model follows the "Dutch" wind scheme, i.e. \citet{Vink+2000} during its long-lasting main sequence phase (MS) in the beginning of its evolution (for $T_{\rm eff} > 10^4{\rm K}$) and \citet{de-Jager+1988} for its cool, RSG phase.  As we see in Figure~\ref{fig:progenitor}, the mass loss rate slightly increases during its early MS, but this variation is not significant for our modeling so for simplicity we choose a constant average value of $\sim 1.2 \times 10^{-8}~{\rm M_{\odot}~yr^{-1}}$ for the first 12.4 Myrs (depicted by a horizontal blue line). As the wind velocity is also an important ingredient of our hydrodynamical  simulation, we assume typical wind velocities based on the star's surface escape velocity. For the early MS phase we find a typical escape velocity of $v_{\rm esc} = 900 \rm km~s^{-1}$ (blue line at the bottom left panel of Figure~\ref{fig:progenitor}) which we multiply with a factor of 2.6 for the assumed terminal wind velocity \citep{Lamers+1995, Vink+2000}, resulting in $v_{\rm wind} = 2340~\rm km~s^{-1}$.

As the star evolves during its late MS, and its surface temperature drops below $T_{\rm eff}\sim 25,000 K$, it experiences a sudden increase of the mass loss rate to $\sim 3 \times 10^{-7}~{\rm M_{\odot}~yr^{-1}}$ (orange line), due to the extra iron opacity, called the bi-stability jump \citep{Vink+2000}. Simultaneously, its escape velocity decreases to $v_{\rm esc} \sim 600 \rm km~s^{-1}$ during this phase that lasts for 1.6 Myrs, due to its radial expansion. Assuming a lower multiplication factor than before of 1.3 \citep{Lamers+1995, Vink+2000}  leads to $v_{\rm wind} = 780~\rm km~s^{-1}$. 

After core hydrogen exhaustion, the massive star expands during its fast thermal timescale, becoming a RSG. The mass loss rate especially during this phase is highly uncertain \citep[e.g.,][]{Smith2014}, but is found empirically \citep[e.g.,][]{van-Loon+2005,Beasor+2020, Yang+2023} and even theoretically \citep{Kee+2021}, to abruptly increase by orders of magnitudes. In our case, we find an average $\sim 6 \times 10^{-6}~ \rm {M_{\odot}~yr^{-1}} $ for the remaining 0.9 Myrs of the star's life (red line), until its eventual Type II SN explosion. Its radial expansion and decrease in mass results in a typical $v_{\rm esc}\sim 70~\rm km~s^{-1}$ and we assume a wind velocity of 20 $\rm km~s^{-1}$, following a multiplication factor of $0.30$  \citep[similarly to][]{Hurley+2002, Belczynski+2008}.

In the right plot of Fig. \ref{fig:progenitor} is depicted the circumstellar structure sculptured by the three phases of wind activity described above and as extracted by our hydrodynamic simulation. The resulted CSM consists of an extended cavity with radius $R_{cav.} \sim 36 \times 10^{18}$~cm ($\sim 12$~pc)  excavated by the early MS and the subsequent late MS wind phase, and forming a density wall of swept up ISM at the outer boundary of the CSM. The inner cavity has been partially filled by a dense wind bubble formed by the RSG wind  ($R_{\rm RSG,bubble} \sim 18 \times 10^{18}$~cm). 
 
\subsection{The evolution of the supernova remnant}

Figure \ref{fig:MMSNR} illustrates the 2D  maps of the gas density, specific bolometric luminosity (luminosity per unit volume), temperature, and velocity  for a sequence of snapshots of the  SNR evolution within the previously formed  CSM. The overall properties that the SNR displays pass through several distinctive phases.  The most characteristic ones being the following: {\bf (a):} At the initial phase of the SNR evolution (Fig. \ref{fig:MMSNR}a; $t= 800$ yrs), the remnant is expanding within the RSG bubble revealing a typical shell-type morphology, consisting of two hot and bright shells of shocked CSM and ejecta gas that have been compressed between the SNR forward and reverse shock, respectively. {\bf (b):} At about $2,500$~yrs after the  explosion (Fig.  \ref{fig:MMSNR}b), the SNR forward shock has penetrated the RSG bubble and evolves within the cavity of the MS wind. At the same time,  a dense shell of shocked ejecta has been formed, due to the action of the reverse shock, lying behind the remnant's contact discontinuity.  {\bf (c):} After $t= 3,500$ yrs of evolution, the SNR has collided with the density wall of the MS wind bubble (Fig. \ref{fig:MMSNR}c). Consequently, the SNR's forward shock gets substantially decelerated as it starts to penetrate the dense shell of the MS wind bubble. Simultaneously, the collision of the SNR with the density wall triggers a strong reflected shock that rapidly moves inwards reshocking the CSM and ejecta material. {\bf (d):} The reflected shock reaches and reshocks the dense shocked ejecta shell lying behind the SNR contact discontinuity (Fig.\ref{fig:MMSNR}d; $t= 4,500$ yrs). At this phase the SNR displays two  distinctive and concentric bright shells  of shocked, hot gas. {\bf (e):} At about 7,500~yrs after the explosion  the reflected shock reaches the center of the SNR (Fig.\ref{fig:MMSNR}e). The whole CSM and ejecta material has been reshocked to high temperatures ($T~>~10^7$~K) and it moves inwards being accumulated at the central regions of the remnant. Immense Rayleigh–Taylor instabilities are formed during this phase, occupying a large fraction of the SNR.  {\bf (f)-(g):} The dense ejecta shell gets assembled at the central regions of the SNR and it gets decelerated. This deceleration of the supersonically moving plasma is communicated to the freely infalling ejecta by the reflected shock that bounces back from the center of the SNR and moves outwards reshocking and heating the accumulated  material (Fig. \ref{fig:MMSNR}f; $t= 8,500$~yrs and Fig. \ref{fig:MMSNR}g; $t= 10,000$~yrs).  Consequently, at this state the SNR displays at its central regions a hot ( $T> 10^7$~K), dense, and bright plasma, rich in ejecta material.  Simultaneously,  the SNR forward shock has reached deeper layers of the MS density wall and it has been substantially slowed down  ($u < 100~\rm km s^{-1}$).  The temperature of the shocked gas behind the forward shock has dropped below $10^4$~K. {\bf (h):} Finally, the shocked ejecta expands, its density drops and its specific luminosity decreases by at least an order to magnitude (Fig. \ref{fig:MMSNR}h;  t= 12,000~yrs). 

This physical process will be repeated again:  the expanding ejecta material will collide with the density wall of the wind cavity, triggering a reflected shock that will move inwards and reshocking the CSM and ejecta material. Nevertheless, in this second round the reflected shock is not strong enough while the CSM and ejecta material has been almost uniformly distributed all over the region of the remnant. Thus, from now on, no  essential central emission is present anymore.   

\subsection{On the emission properties of the supernova remnant}
 To clarify further the emission properties of the resulting SNR, in Figure \ref{fig:Xray_opt} it is illustrated the specific luminosity of the shocked gas whose temperature is below $T= 10^4$~K (left semicircle of each plot) as compared to those whose gas temperature is higher than  $T= 10^7$~K (right semicircles). These 2D  maps work as an indicator of the SNR regions the emission of which is dominated by thermal X-ray photons ($T > 10^7$~K)   and of those where the shocked gas is cold enough to a level where optical line emission is anticipated ($T < 10^4$~K).   At the initial phase ($t= 800$~yrs), where the SNR is young and well-within the dense RSG bubble, the remnant displays a typical shell-type morphology of two X-ray bright shells. In contrary, when the SNR forward shock penetrates the wind bubble and starts to evolve into the MS cavity  ($t=2,500$~yrs) the luminosity of the post-shock gas drops substantially due to the low density of this region. Thus, in this phase the remnant possesses one bright X-ray shell corresponding to the dense shocked, ejecta gas lying behind the reverse shock. At the moment of the collision of the SNR with the MS density wall ($t=3,500$~yrs), the forward shock re-brights again as it starts to penetrate the dense wind/ISM shell. At that moment its velocity remains high enough and thus, the post-shock gas emission in dominantly in the X-ray band.   The shell behind the reverse shock remains X-ray bright but its emission gradually faints due to its expansion.   Nevertheless, when the reflected shock reaches the shocked ejecta shell ($t=4,500$~yrs), its specific luminosity increases by at least an order of magnitude. Consequently, at this state the remnant displays two concentric X-ray bright shells, with the inner one possessing the higher temperature (compare with Fig \ref{fig:MMSNR}d).  It is intriguing that such a phase shares very similar characteristics with what has been observed for the case of the MMSNR Kesteven~27 \citep{Chen2008} where the remnant reveals two distinctive and concentric shell with the inner one being the hotter.  

When the reflected shock reaches the center of the remnant and subsequently bounces back ($t=7,500$~yrs, 8,500~yrs and 10,000~yrs), the re-shocked gas has been accumulated at the center of the remnant, being heated to temperatures well above  $10^7$~K. At this phase the SNR hosts a hot, dense, ejecta-rich and bright X-ray emitting gas at its center. At the same time, a very thin and bright shell has been formed just behind the SNR’s forward shock. The remnant’s blast wave has penetrated deeper layers of the wind’s density wall and thus, it has been substantially decelerated ($ u< 100~\rm km~s^{-1}$). The high density post-shock gas is characterized by high specific luminosities, while its temperature has dropped below the  $10^4$~K. Under such conditions, the remnant’s forward shock in not any more adiabatic but instead it resembles an optically bright, radiative shock. Due to the non-negligible energy losses, the shocked shell has been collapsed into a very thin filamentary structure. Overall,  during this time interval ($t \sim 7,500 - 10,000$~yrs), the SNR has been transformed into a typical MMSNR revealing a centrally peaked, thermal X-ray emission surrounded by a slow, radiative and filamentary structured blast wave which thermal radiation is expected to be dominated by optically line emission. These characteristics are met to several MMSNRs such as HB3, VRO 42.05.01 and W44 \cite[e.g.][respectively]{Boumis2022, Derlopa2020, Mavromatakis2003}   which possess extended outer filaments bright in  optical emission lines, coexisting with a centrally peaked thermal X-ray emission. 

Finally, the central X-ray emission of the remnant gradually faints out as the shocked ejecta expands and its temperature and density decreases ($t= 10,000$~yrs). Hence, the SNR losses its mixed-morphology properties and evolves as a `classic’ mature optically bright remnant.

\begin{figure*}
\includegraphics[trim=0 610 0 35 ,clip=true,width=\linewidth,angle=0]{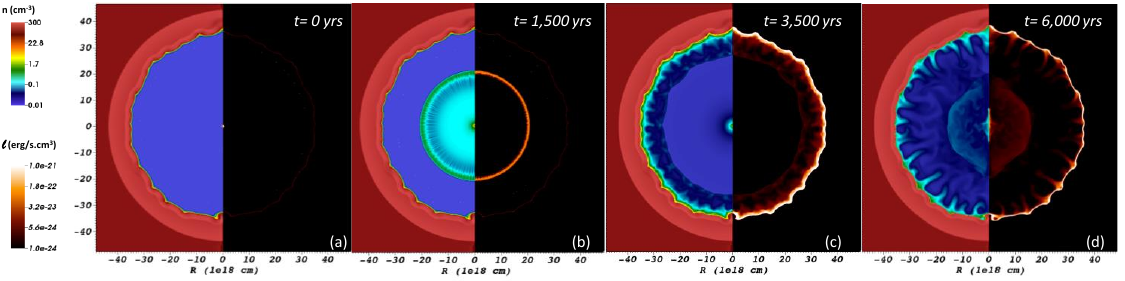}
\caption {The 2D density and specific luminosity  maps of a SNR evolving within the wind-blown cavity of the $15~\rm M_{\odot}$ progenitor star but without taking into account the RSG wind phase (see text for details).   }
\label{fig:MMSNR_noRSG}
\end{figure*}

\section{Discussion and summary} \label{Sect5}

In this work by coupling the predictions of  stellar evolution theory  with hydrodynamic simulations, we modeled the evolution of a SNR by a 15~$\rm M_{\odot}$ stellar progenitor  and its interaction with the surrounding CSM formed by the parent star's  mass outflows during the pre-explosion phases. We showed that the mixed-morphology properties observed in several SNRs can naturally arise by the action of a strong reflected shock  that assembles and reshocks the SN ejecta at the central regions of the remnant.

In particular, due to the interaction of the SN ejecta with the complex surrounding CSM, the resulting SNR's overall morphology passes through several distinctive stages: starting from a bright, young, shell-type SNR and then followed by a phase where it exposes two bright, concentric X-ray inner shells,  the remnant is then transformed into a typical MMSNR possessing centrally peaked, X-ray thermal emission. Finally, the remnant ends its evolution as a mature shell-type SNR consisting of slow, radiative shocks.  In this aspect, the mixed morphology properties of a SNR originating from a RSG progenitor represents a specific evolutionary feature of the remnant, while the state of the double concentric X-ray shells observed for the case of Kesteven~27 \citep{Chen2008} represents an antecedent phase before the centrally peaked X-ray SNR emission.

Does this mean that all remnants of Type II SNe will display -for a given time window of their lifetime- MMSNR properties? Clearly not, as the key ingredient for the MMSNR transformation is the local ISM density. High ambient medium densities are required in order to constrain the MS wind bubble close to the explosion center and to form dense cavity walls that are necessary to trigger a strong reflected shock \citep{Sgro1975, Dwarkadas2007}. In a different case -e.g. for typical warm/hot ISM densities- the SN ejecta will be distributed all over the formed large  cavity and thus, no sufficiently dense material will be met by the reflected shock  capable of producing the required central X-ray emission. This evidence is alighted to the fact that MMSNRs are associated with denser parts of the ISM \citep{Vink2012}.

Even if our work did not aim to model a specific MMSNR -as such a task requires     detailed fine-tuning of the SN and CSM/ISM properties - and it was limited to one typical study case (a $15~\rm M_{\odot}$ single, non-rotating progenitor star within an ISM of $n= 100~cm^{-3}$), several features observed in a number of MMSNRs where reproduced. The time interval of our models that was required for the action of the reflected shock to assemble the SN ejecta/CSM material at the center of the remnant and ignite the central X-ray emission ($t \sim 7,500 -10,000$~yrs) is comparable to the estimated ages of the majority of MMSNRs \citep[$t_{\rm MMSNRs} \approx 10^4$~yrs,][]{Vink2012}.  Regarding the properties of the X-ray emitting gas, our models reproduce the approximately homogeneous temperature of $T \sim 10^7 - 10^8$~K observed in MMSNRs, while the fact that the central emitting gas in our hydrosimulations consists mostly by ejecta material is aligned with the enhanced abundances and the metal-enrichment that the X-ray spectroscopy reveals for a number of MMSNRs \citep[e.g.][]{ Lazendic2006, Pannuti2014}.  As for the outer SNR shell emission, our models show that during the MMSNR phase, the remnant’s forward shock carries the characteristics of a slow, radiative shock. The cold ($T<10^4$~K) and dense ($n> 200~\rm cm^{-3}$) post-shock gas has collapsed into a filamentary shell due to intense radiation losses and the short cooling timescales that govern this region. Such morphological and thermodynamical conditions are extracted by optical observations conducted for a number of MMSNRs \citep[e.g.][]{Fesen1995, Pannuti2017, Boumis2022}.

Finally, as far as  the radio, non-thermal synchrotron emission of the outer shell is concerned -a characteristic feature met in all MMSNRs- we cannot extract direct conclusions form our modeling, as for this it is required magnetohydrodynamic simulations and additional assumptions on the strength and the orientation of the ambient magnetic field.  Nevertheless, according to our modeling,  the time interval during which the SNR possesses its centrally peaked X-ray emission ($t_{\rm SNR} \sim 7,500-10,000 $~yrs), the forward shock is well within the wind blown cavity wall and its velocity has been dropped bellow  the 200 $\rm km~s^{-1}$. For such low shock velocities, the bulk of the syncrotron emission is expected to be in the radio band \citep[e.g.][]{Reynolds2008, Vink2012}. In addition, the propagation of SNR's blast wave into the dense medium of the swept-up wind/ISM shell is expected to cause a substantial increase of the synchrotron emission luminosity, as the number of the accelerated electrons is generally found to be a fraction of the total particle density \citep[e.g][]{Ellison2005}.  The fact that the outer shock is radiative consists an additional factor towards the enhancement of the outer shock's synchrotron emission  as the shock's compression factors are large giving rise to strongly compressed magnetic fields and higher cosmic-ray electron densities \citep[see][]{Cox1999}.

 Comparing our results to those of \citet{Dwarkadas2013} -as both papers deal with the idea of a reflected shock triggered by the progenitor star's wind-blown cavity- it is clear that the final outcomes and conclusions deviate substantially. In particular, \citet{Dwarkadas2013} found no essential central X-ray emission in their models as the central regions of the SNRs were described by a very low density ejecta gas. The key difference between our model and theirs is the existence of the dense RSG bubble at the center of the CSM structure.  The  inner RSG wind bubble plays a vital role at the evolution of the SNR and the formation of its mixed morphology properties. Particularly, the dense wind bubble decelerates the SNR at the early phases of its evolution triggering an active reverse shock that sweeps up the freely expanding ejecta and forms a dense  ejecta shell. This dense ejecta shell is well maintained during the whole SNR evolution and consists of the material that will be shocked and produce the central X-ray emission after the action of the reflected shock.  To verify this statement, we re-run our simulations but this time without the RSG phase (i.e. we introduced the SN ejecta right after the end of the late MS  phase). The results of our simulations (see Fig. \ref{fig:MMSNR_noRSG}) show that even if a similar process occurs -i.e. a strong reflected shock is formed that reheats the SN material at the center of the remnant- due to the low density of the expanded ejecta, the produced X-ray luminosity is  one to two orders of magnitude lower than our models' that involve the RSG bubble, in agreement with the results  of \citet{Dwarkadas2013}. 

We refrain a direct comparison between our work and other MMSNRs models of the literature -that included reflected shocks- which have been dedicated to specific SNRs and thus, their simulations were set up on the basis of observational results and not on specific evolutionary progenitor’s paths \citep[e.g.][ see Introduction]{Zhou2011, Ustamujic2021}.   Nevertheless, it is intriguing to be investigated whether the ‘tailor-made’ circumstellar structures of these models i.e. the ring-like CSM for W49B  \citep{Zhou2011} and the toroidal cloud for IC 443 \citep{Ustamujic2021} -demanded by the models in order to reproduce the relevant observables - could be reproduced by  the post-main-sequence mass loss ejected from the given progenitor (either in the form of equatorial confined RSG stellar winds or episodic mass loss). 

This work was focused on a 15~$\rm M_{\odot}$  progenitor star. Nevertheless, our main  conclusions  can be extended to different masses of RSG progenitors/Type II SNe. Our results are not expected to be sensitive to the exact values of wind mass-loss rates and velocities, as the wind mechanical luminosity ($L= \frac{1}{2}\dot{M}u_w^2$) during the MS phase can be counterbalanced by different ISM densities resulting to comparable sizes of the wind-blown cavities \citep{Weaver1977}. Depending on the mass of the progenitor star and the local ISM properties, different outcomes are expected regarding the brightness of the central X-ray emission and the time intervals that the remnants will possess mixed morphology properties.

Furthermore, it is tempting to speculate whether other SNe progenitor paths can reproduce the key ingredients of our model required for the formation of a MMSNR,  namely an extended circumstellar cavity being partially filled with an inner dense bubble. It is unclear whether mechanisms of much higher mass loss rate (of a few orders of magnitude higher than the RSG phase assumed in this study) but lasting a much shorter timescale (only a few months to centuries before the explosion itself) could lead to similar conditions for the  formation of a MMSNR phase. These mechanisms could potentially involve a sudden increase in mass loss or an ``outburst'' ejection  before the core-collapse of a massive star \citep[e.g.,][]{Davies+2022}, potentially due to instabilities in the late burning phases of the stellar progenitor \citep[e.g.,][]{Smith+Arnett2014, Fuller2017,  Wu+Fuller2021, Linial+2021}, or due to a unstable binary mass transfer and a subsequent common envelope ejection well-timed with the SN itself \citep[e.g.,][]{Mcley+Soker2014}, or even due to pair-pulsations  \citep[e.g.,][]{Renzo+2020} before a successful SN explosion. It would be interesting also to consider SN events that transition from Type I to II due to a previously ejected H-rich shell located around the progenitor \citep[e.g., SN2014C;][]{Margutti+2017, Brethauer+2022} in the context of MMSNRs, and whether there  could be a link between them. Finally, similar speculations can be done in the framework of Type Ia SNe, as e.g. is the progenitor of the MMSNR W49B \citep{Zhou2018}.  The low-mass star progenitors of Type Ia SNe  are not able to form the extended wind-blown cavities met around OB stars. However, mass outflows capable of excavating the required cavities are predicted to emanate from the surface of the white dwarf-progenitor either in the form of `accretion winds'  \citep{Hachisu1999, Badenes2007} or during the contraction of the stellar core towards the formation of a white dwarf during the post-AGB phase \citep{Paczynski1971, Chiotellis2020}. The dense inner wind bubble needed for the formation of MMSNRs could be formed by a giant donor star (symbiotc channel)  or by an ejected common envelope in the single and double/core degenerate regime, respectively \citep[e.g][]{Hamuy2003, Dilday2012, Chiotellis2012, Broersen2014, Soker2015, Meng2017}.

We aim for further investigation of RSG progenitors of different initial masses and assumed stellar physics, surrounded by ISM  of varying densities as well as to assess the alternative evolutionary paths mentioned above that may lead to MMSNRs. This is needed to constrain the required conditions for the occurrence of the MMSNR phenomenon, their life duration, and as well as their formation rate.  In addition, a detailed and fine-tuned modeling of well known MMSNRs within the framework of our model is required, where the quantitative comparison between the simulations and the relevant observables, will highlight in which cases and to what extent the reflected shock model proposed in this work is capable to explain the nature and evolution of this peculiar class of SNRs.

\section*{Acknowledgements}

We thank the anonymous referee for their  feedback that helped to improve the paper. AC gratefully acknowledges the organisers of the Lorentz Center's Workshop `Supernova Remnants in Complex Environments' (Leiden, Netherlands) for organising such an inspiring workshop and the participants for the intriguing discussions on the topic of MMSNRs. Special thanks to A. Filopoulou, Prof.~E. Pleionis and A. Dimou for all the support and motivation. EZ acknowledges support from the Hellenic Foundation for Research and Innovation (H.F.R.I.) under the “3rd Call for H.F.R.I. Research Projects to support Post-Doctoral Researchers” (Project No: 7933).  This work has been supported by the grant PID2021-124581OB-I00 funded by MCIN/AEI/10.13039/501100011033 and 2021SGR00426 of the Generalitat de Catalunya. This work was also supported by the Spanish program  Unidad de Excelencia Mar\' ia de Maeztu CEX2020-001058-M.


\section*{Data Availability}

The data underlying this article will be shared on reasonable request to the corresponding 
author.



\bibliographystyle{mnras}
\bibliography{MMSNR}







\bsp	
\label{lastpage}
\end{document}